\documentclass[%
aip,jcp,amsmath,amssymb, reprint,twocolumn
]{revtex4-2}
\usepackage{amsmath}
\usepackage{mathtools}
\usepackage{amsfonts}
\usepackage{amssymb}
\usepackage{dcolumn} 
\usepackage{array}
\newcolumntype{P}[1]{>{\centering\arraybackslash}p{#1}}
\newcolumntype{M}[1]{>{\centering\arraybackslash}m{#1}}
\newcolumntype{C}[1]{>{\centering\arraybackslash}p{#1}}
\usepackage{float}
\usepackage{psfrag}
\usepackage{graphicx}
\usepackage{diagbox}
\usepackage[caption=false]{subfig}
\usepackage{tikz}
\usetikzlibrary{positioning}
\usetikzlibrary{arrows}
\usetikzlibrary{trees}
\usetikzlibrary{decorations.pathmorphing}
\usetikzlibrary{decorations.markings}
\usetikzlibrary{automata,positioning}
\usepackage{braket}
\usepackage{rotating}
\usepackage{adjustbox}
\usepackage{ragged2e}
\usepackage{simplewick}
\usepackage{simpler-wick}
\usepackage{csquotes}
\usepackage{xcolor}
\usepackage{fancyhdr}
\usepackage{csquotes}
\UseRawInputEncoding
\begin{document}

\author{Valay Agarawal}
\affiliation{ Department of Chemistry, Indian Institute of Technology Bombay}
\author{Chayan Patra}
\affiliation{ Department of Chemistry, Indian Institute of Technology Bombay}
\author{Rahul Maitra}
\email{rmaitra@chem.iitb.ac.in}
\affiliation{ Department of Chemistry, Indian Institute of Technology Bombay}

\title{An Approximate Coupled Cluster Theory via Nonlinear Dynamics and Synergetics: the Adiabatic Decoupling Conditions}

\begin{abstract}
The coupled cluster iteration scheme is analysed as a 
multivariate discrete-time map using nonlinear 
dynamics and synergetics. The nonlinearly 
coupled set of equations to determine the cluster 
amplitudes are driven by a fraction of the entire 
set of the cluster amplitudes. These driver amplitudes 
enslave all other amplitudes through a synergistic 
inter-relationship, where the latter class of amplitudes 
behave as the auxiliary variables. The driver and the 
auxiliary variables exhibit vastly different time scales 
of relaxation during the iteration process 
to reach the fixed points. The fast varying
auxiliary amplitudes are small in magnitude, while the 
driver amplitudes are large, and they have a
much longer time scale of relaxation. Exploiting 
their difference in relaxation time-scale, we
employ an adiabatic decoupling approximation, where 
each of the fast relaxing auxiliary modes are expressed 
as unique functional of the principal amplitudes. This
results in a tremendous reduction in the independent 
degrees of freedom. On the other hand, 
only the driver amplitudes are determined accurately via
exact coupled cluster equations. We will demonstrate
that the iteration scheme has an order of magnitude 
reduction in
computational scaling than the conventional scheme. With 
a few pilot numerical examples, we would demonstrate that
this scheme can achieve very high accuracy with significant
savings in computational time.
\end{abstract}

\maketitle

\section{Introduction}
Coupled Cluster (CC)\cite{cc3,cc4,cc5,bartlett2007coupled} 
is a versatile tool for accurately solving electronic
Schr{\"o}dinger equation. In single reference CC (SRCC) 
method, one introduces an exponential wave-operator 
$\Omega$ which induces electronic correlation by its action
on a reference determinant: 
$|\Psi_{CC}\rangle = \exp(T)|\Phi_0\rangle$. Here 
$T$ is a sum of many-body hole-particle excitation 
operators and $|\Phi_0\rangle$ is the reference function,
which, in closed shell SRCC, is usually chosen to be
Hartree-Fock (HF) determinant. The cluster operators, which
are the unknown variables, are determined by projecting 
the similarity transformed effective Hamiltonian 
$H_{eff}=e^{-T}He^T$ against the excited determinants. 
Thus the effect of
excited determinants is folded on to the reference 
determinant, and one may compute energy by evaluating
the expectation value of the effective Hamiltonian 
$H_{eff}$ with respect to the reference determinant
$|\Phi_0\rangle$: $E_{corr} = \langle H_{eff} \rangle = \langle e^{-T}He^T \rangle$.

Due to the exponential parametrization of the wave operator,
the equations to determine the cluster amplitudes are 
inherently nonlinear. Thus the usual way to solve for the
amplitudes is to employ an iterative procedure, and 
demand that each amplitude corresponding to the 
$H_{eff}$ matrix element
between the ground and the excited states is zero upon
convergence of the iterative procedure. In other words,
$h_\mu=0$, where $h$ is the amplitude associated with 
$H_{eff}$ and $\mu$ denotes the combined hole and particle 
labels associated with the excitations. This implies that
for the CC theory with singles and doubles excitations
scheme, one needs to solve for ($n_on_v+n_o^2n_v^2$)
number of cluster amplitudes, where $n_o$ and $n_v$ are 
the numbers of the hole and particle orbitals respectively.
Note that in CC theory, all the cluster operators are 
nonlinearly coupled. Owing to the hole-particle excitation
structure, the cluster operators do not
contract among themselves; however, various powers of the
cluster operators necessarily contract with the one and
two-body terms in Hamiltonian. This ensures that the 
effective Hamiltonian may contain up to quartic power
of the cluster operators in the CCSD scheme. Due to the
contraction of the cluster operators with the Hamiltonian, 
the method scales as $n_o^2 n_v^4$ at worst, which is a 
formidable scaling for large molecular systems. 

There have been several manuscripts addressing the steep 
computational scaling of CC theory, and prescribing
different ways to circumvent it. However, little effort has
been made to approach this issue by exploiting the inherent
nonlinearity of the amplitude equations. This manuscript 
aims to provide a recipe to address this problem, and 
provide a solution by analysing the time series associated
with its iteration scheme and exploiting an interdependence
of the amplitudes which is usually evident in multivariate
dynamics. Szak{\'a}cs and 
Surj{\'a}n\cite{szakacs2008iterative,szakacs2008stability}, 
for the first time, demonstrated that the CC iteration 
scheme under an input perturbation shows chaotic dynamics.
Some of the present authors had shown that the dynamics 
associated with a double similarity transformed CC\cite{maitra_correlation_2017,maitra_coupled_2017,tribedi2020formulation}  iterations follows the universality of time-discrete
map of one-parameter family. The authors analysed the 
phase space trajectory under an input perturbation and 
demonstrated a full period-doubling bifurcation cascade 
that precedes the onset of the chaos. As such, it was shown 
that such a time-discrete multivariate map of one parameter
family obeys the universality of Feigenbaum dynamics\cite{feigenbaum1978quantitative}. It 
was also established that there exists interesting
inter-relationship among the cluster operators. In a
follow-up work\cite{agarawal2020stability}, the 
authors had demonstrated that the macroscopic pattern of 
the entire iteration dynamics is governed by a few
\textit{principal} cluster operators, which were
taken to be the unstable modes, and they were assigned 
to be the order parameters.
On the other hand, there exist numerous amplitudes which
are smaller in magnitude, and they are enslaved under the
principal amplitudes. They were described as the 
\textit{auxiliary} amplitudes. As such, it was 
conjectured that 
the auxiliary amplitudes can be expressed as unique 
functionals of the principal amplitudes. Based on this
observation from non-linear dynamics and Synergetics
\cite{Haken_1989, haken1982slaving, Haken1983}, the 
authors introduced a supervised shallow
machine learning (ML) model, based on Kernel Ridge 
regression
\cite{scikit-learn} technique, to train the auxiliary amplitudes \cite{scikit-learn}
as functions of the principal amplitudes. The 
resulting hybrid CC-ML scheme was demonstrated to 
have remarkable sub-microHartree accuracy compared to 
the canonical CCSD scheme with 40-50\% savings of 
the computational time\cite{agarawal2021accelerating}. 

In this work, we analytically establish the 
inter-relationship among the cluster operators during 
the iteration process. We analyse the relaxation 
time for each of the cluster operators to reach their 
predefined converged values, which are taken to be the 
fixed points of the dynamics. We establish that the 
principal amplitudes, which are large in magnitude take 
a higher number of steps to reach their fixed points,
while the smaller
auxiliary amplitudes take much fewer iterations. As such, 
there is a significant difference in which these two sets 
of amplitudes relax, although there is no distinct 
boundary to 
define these two sets of amplitudes. We have exploited this
different time scales of relaxation to decouple their
equations; In what follows, we will show that the under 
an \enquote{adiabatic} approximation, the auxiliary amplitudes
can be written analytically as functions of the 
principal amplitudes alone. Contrary to that, the principal
amplitudes, which are much fewer in numbers, 
can be accurately 
determined via a feedback coupling of two sets of 
amplitudes. This amounts to assigning the principal 
amplitudes as the independent variables of the dynamics
while the auxiliary amplitudes are the dependent 
variables. Through this decoupling procedure, we would 
demonstrate that there is a tremendous reduction in the 
independent degrees of freedom in the iteration dynamics.
This results in a significant reduction in the computational
scaling, which we will demonstrate analytically and 
numerically.

The paper is organised as follows: in Section II, we 
establish the concept of principal and auxiliary 
amplitudes, and show that they have different time scales
to reach their respective fixed point solutions. This would
motivate us to express the dependent auxiliary amplitudes 
as the functions of the principal amplitudes. Exploiting 
this, we would outline a bird's-eye-view of the newly
developed method in Section III which we would refer 
as the Adiabatically Decoupled CC (AD-CC) theory. In 
Section IIIA, we explicitly 
present the adiabatic decoupling conditions where 
we analytically show the dependence of the auxiliary
amplitudes on the principal amplitudes, and derive the
working equation to 
determine the auxiliary amplitudes. Section IIIB deals
with the feedback coupling mechanism where the principal
and the auxiliary amplitudes are allowed to couple to 
determine the former set of amplitudes. We would demonstrate
the genesis of two non-equivalent schemes, and 
analyse the computational scaling associated each of them. In Section IV, we demonstrate the efficacy of the
newly developed schemes within CC doubles (CCD)
approximation by applying it to a number of molecular
systems with varied electronic complexities. In this 
section, we would also numerically analyse the scaling of the methods
by taking a few molecular examples. Finally, in Section V,
we will summarise and present a few probable future 
directions.


\section{The Principal and the Auxiliary Amplitudes, and Their Relaxation Time Scale:}
\begin{figure}
    \centering
    \includegraphics[width=\linewidth]{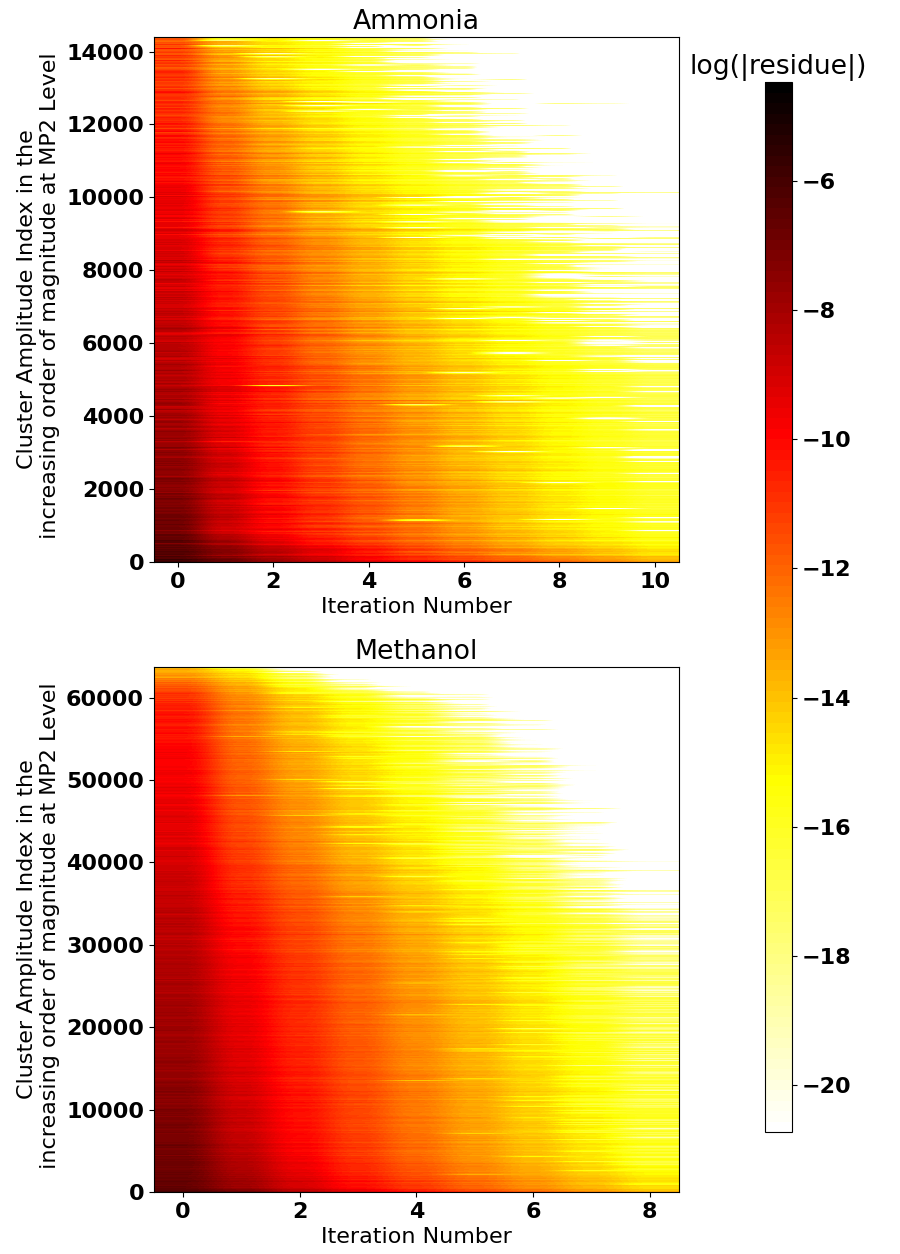}
    \caption{The logarithm of the absolute value of each of the residues (above $1 \times 10^{-9}$) at different iterations for Ammonia and Methanol in cc-pVDZ basis. Note that the first order amplitudes are arranged in ascending order top to bottom in the vertical axis. The larger the amplitudes (as one moves down from top), the larger is the horizontal spread of the red-yellow portion, implying larger variation and longer time scale of relaxation to reach the fixed point solutions.}
    \label{fig:variance}
\end{figure}

As we mentioned previously in the introduction, the
different cluster amplitudes have different time scales 
in reaching their fixed point solutions, which are taken 
as the converged values of the 
amplitudes with a predefined tolerance. In order to show
this difference in the relaxation time-scale for 
different amplitudes, in Fig. \ref{fig:variance}, we 
have shown the
variation of the absolute residues (defined as 
the numerator on the right-hand side in Eq. \ref{eq1}) 
as a function of the iteration step for two different
molecules. Note that the residue is a measure of the
difference of the amplitudes in successive iterations 
(see Eq. \ref{eq1}). Here in the vertical axis, we have
arranged the first order cluster amplitudes in an 
increasing order of magnitudes top to bottom, and this 
is expected to give a fairly good estimate of their 
relative ordering even after the convergence. 
Here in the plot, the red colour 
signifies the larger magnitude of the residue, while
the smaller ones are towards white. Note that 
larger the magnitude of the amplitudes, the larger
is the variation of the amplitudes (i.e., a 
larger magnitude of the residues). Also, note that
the larger amplitudes take a longer time to settle. 
This implies that the residues associated to these larger
amplitudes take much higher number of iterations to go
below a threshold of tolerance. They are thus the unstable 
modes of the dynamics. On the other hand,
smaller the amplitudes, the residues take fewer iterations 
to get below the threshold, and thus they are the stable 
modes in the dynamics. Clearly, there is a difference 
in the time scale for the large and the small amplitudes
to reach their respective fixed points. 

In this article, we exploit this difference in their 
time scale to simplify the convergence procedure. In what
follows, we would show that it is possible 
to restrict the iteration process to determine
only the largest amplitudes which have longer relaxation
time scale, while the smaller amplitudes would simply 
be parametrized as the functionals of the previously
mentioned large amplitudes. This is 
the central topic in Synergetics where one deals with the 
dimensionality reduction in multivariate nonlinear
dynamics to work with fewer degrees of freedom.
At this stage,
let us set the nomenclatures and notations. The large
amplitudes that play the the role of unstable modes, and 
drive the overall iteration process will be
termed interchangeably as the \textit{driver} or the
\textit{principal} amplitudes. Space spanned by these
principal amplitudes will be termed as the 
largest subspace (LS), and the operators
corresponding to the principal amplitudes will 
simply be called as the principal cluster operators. 
Let us generically denote these principal amplitudes as 
$t_L$. Thus the set $\{t_L\}$ denotes the collection of 
all the principal cluster amplitudes which belong to the 
LS. Let us denote their dimension as $n_L$. On 
the other hand, the smaller amplitudes (the stable modes 
of the dynamics, having a shorter 
time scale) which supposedly \enquote{work under} the 
principal amplitudes would be referred to as the 
\textit{auxiliary} amplitudes. In Synergetics, the 
principal and auxiliary amplitudes are referred to as 
the \enquote*{master} and \enquote*{slave} variables. The space spanned
by the slaved amplitudes will be termed as the 
smaller subspace (SS). Let us denote the set of 
the auxiliary amplitudes generically as $\{t_S\}$. 
The $S$ in the subscript stands for \enquote*{small} to indicate 
that they are much smaller in magnitude than $t_L$'s.
The dimension of the smaller subset, denoted by $n_S$ is
much larger than $n_L$: $n_S >> n_L$. Note that at this 
stage, we have not followed any established procedure 
to distinguish the principal and the auxiliary amplitudes. 
Rather, the principal amplitudes are taken to be a small
fraction (10\%-20\% of elements having the highest
magnitude) of the total nonzero amplitudes. 
However, a more rigorous choice of the principal amplitudes 
and the LS can be made based on the information 
theoretical techniques, or by analysing the Lyapunov 
exponents
arising out of the Jacobian stability matrix. This will 
be the topic of a future publication. In the current 
work, by exploiting the difference in the relaxation time
scale of the principal and auxiliary amplitudes, we will
analytically establish their interdependence. We would
also schematically show how an iteration scheme can be built
at a much cheaper computational scaling based on the 
above observations.
\section{The Circular Causality Relationship between the Principal and Auxiliary Amplitudes:}
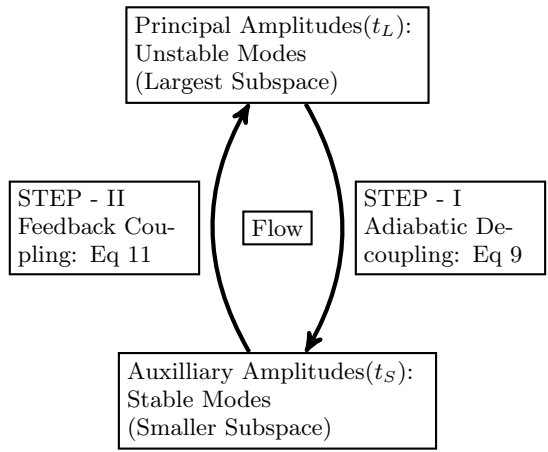
\begin{figure}[!h]
    \centering
\begin{tikzpicture}
  [
    -,
    >=stealth',
    auto,node distance=2.3cm,
    thick,
    main node/.style={rectangle, draw}
    ]
\node[main node]      (maintopic)                              {Flow};
\node[main node]        (LS)       [above of= maintopic,text width=3.8cm] {Principal Amplitudes($t_L$):\\ Unstable Modes\\ (Largest Subspace)};
\node[main node] (ML) [right of= maintopic,text width=2.3cm] {STEP - I \\Adiabatic Decoupling: Eq \ref{eq9}};
\node[main node] (FT) [below of= maintopic,text width=4cm] {Auxilliary Amplitudes($t_S$):\\ Stable Modes \\(Smaller Subspace)};
\node[main node] (CCT) [left of= maintopic,text width=2.3cm] {STEP - II \\ Feedback Coupling: Eq \ref{eq:feedback-coupling}};
\draw[->,ultra thick] (FT.120) to[out=120, in=-120] (LS.-120) ;
\draw[->,ultra thick] (LS.-60) to[out=-60, in=60] (FT.60);
\end{tikzpicture}
\caption{Bird's-eye-view of the circular Causality loop where the principal amplitudes are mapped on to the auxiliary amplitudes using the adiabatic decoupling relation (Eq. \ref{eq9}, step-I), and the feedback coupling to obtain updated set of the driver amplitudes via step-II.}
    \label{fig:circ-caus}
\end{figure}

As discussed previously, the fundamental assumptions of 
Synergetics dictate that the macroscopic progression of 
the iteration process is almost entirely governed by the 
principal cluster amplitudes belonging to LS. As shown 
in Fig. \ref{fig:variance}, these principal amplitudes 
have a longer time scale of relaxation to reach the fixed
point solutions. 
Contrary to that, there is the smaller subspace, SS, which
contains all the remaining smaller cluster amplitudes. 
The variation of the auxiliary amplitudes is suppressed and
we can assume that their contribution to the macroscopic
dynamics is asymptotically negligible. These auxiliary 
cluster amplitudes are thus enslaved under the driver
amplitudes, and they are thus often termed as \enquote*{slave}
amplitudes. Following the principles of Synergetics and 
our earlier work, we conjecture that each of the auxiliary
amplitudes can be written as a functional of the principal
driver amplitudes.
\begin{equation}
    t_{S_i}^{\tau} = F_{S_i}(\{t_{L}^\tau\})
\end{equation}
Here small $i$ denotes the index of the auxiliary 
amplitudes and they belong to the smaller subset. 
Thus $i=\{1,2,...,n_S\}$. The set $\{t_L\}$ 
denotes the entire group of the principal
amplitudes which is a collection 
of $n_L$ elements. $\tau$ denotes the 
discrete time step. Assuming that the auxiliary amplitudes 
have a much faster and suppressed variation compared 
to the large principal amplitudes which are characterized
by slow relaxation time scale to reach the fixed points, 
we employ an adiabatic decoupling condition such that one
may approximate the functional form $F$ for each excitation 
$S_i$ associated with the auxiliary amplitudes. Unlike our 
previous attempt to numerically simulate $F$ via 
supervised machine learning, in this work, we analytically 
arrive at a set of equations for each of the auxiliary
amplitudes. This is achieved via decoupling the 
fast and slow relaxing variables, and exploiting the fact
that the variation of the auxiliary amplitudes has little
effect in the macroscopic features of the iteration 
dynamics. Thus their subdynamics is asymptotically
negligible. This allows us to
express each of the auxiliary amplitudes uniquely by 
$t_L$ alone. Contrary to that,
only the principal amplitudes are determined more accurately
via the usual algebraic or diagrammatic techniques which
allows the feedback of the auxiliary amplitudes $t_S$, as 
well as the coupling with other principal amplitudes $t_L$.
Our overall method relies on the adiabatic decoupling of 
the fast relaxing stable modes from the slow unstable 
modes via adiabatic decoupling.
In what follows, we will demonstrate that the determination
of both the principal and auxiliary amplitudes may be 
done with a tremendous reduction in computational scaling
than the conventional solution of the coupled cluster 
equations. We will elaborate upon both these steps in the
subsequent subsections. Here in Fig. \ref{fig:circ-caus}, 
we give a bird's-eye-view 
of the entire iteration scheme which consists of these
two steps connected via what is referred to as the 
circular causality loop.

We first choose the principal cluster amplitudes from the
first order perturbative estimate. At this stage, the choice
of the principal amplitudes is made solely based on the
magnitude of the first order cluster operators, and the
chosen large amplitudes are assumed to span the LS. All 
the remaining 
cluster amplitudes with smaller magnitudes are taken to 
be the auxiliary enslaved variables. While this choice of
the principal amplitudes is good enough for all the
practical purposes, as mentioned before, this is somewhat
ad-hoc and one needs to resort to a more sophisticated
technique to pick up the unstable modes. In an ensuing
article, we would
demonstrate how one may choose the principal amplitudes 
via the maximization of the mutual information.
However, this will not be discussed any further in the
present article. The largest amplitudes are then
mapped on to the smaller auxiliary amplitudes via step-I 
using the adiabatic decoupling condition.
In the next subsection, we will show that in order to
determine the auxiliary amplitudes,
only the principal amplitudes are allowed to contribute, 
while the coupling among the auxiliary amplitudes are 
switched off. Thus each of the 
auxiliary amplitudes are written as unique functional 
of the principal amplitudes. The large driver amplitudes 
and the auxiliary amplitudes thus
obtained via adiabatic decoupling are then allowed to 
couple to obtain an updated set of principal amplitudes 
as shown in step-II. 
This overall feedback mechanism 
is known as the circular causality loop in Synergetics, 
which is continued till all the principal 
amplitudes (and hence the auxiliary amplitudes due to their
fixed functional dependence on the principal amplitudes)
are converged. In the next subsection, we will discuss
both these steps in more detail and show their exact
interdependence. 

\subsection{The Parametrization of the Auxiliary Amplitudes: the Adiabatic Decoupling Condition:}
In this subsection, we would elaborate upon the
forward mapping in which the auxiliary amplitudes
are expressed in terms of the principal
amplitudes, as
shown by step-I in Fig. \ref{fig:circ-caus}.
Here we borrow heavily from the statistical 
description of self-organization where many variables 
coherently \enquote{work together} without any external 
stimuli\cite{Haken1983,[Chapter~7 ]haken2013synergetics}. Let us look at the coupled cluster iteration scheme
as a self-organizing system containing various interacting 
subsystems.

Let us consider the coupled cluster Jacobi iteration scheme: \begin{equation} \label{eq1}
\begin{split}
 \Delta t_{\mu} & = t_{\mu}^{(k+1)} - t_{\mu}^k \\
 & =\frac{1}{D_\mu} (H_\mu + (\contraction{}{H}{}{T} H T)_\mu + \frac{1}{2}(\contraction{}{H}{}{T}
\contraction[2ex]{}{H\;}{T}{}
H\;TT)_\mu + . . . ) 
\end{split}
\end{equation}
Here $T$ is the cluster operator, $H$ is the one and two
electron Hamiltonian matrix element. $t_\mu$ denotes the
associated amplitudes with hole-particle excitation
structure \enquote*{$\mu$}. The notation $(...)_\mu$ 
means the composite inside the parenthesis, generated 
out of the contraction of Hamiltonian and the cluster
operator(s) have the hole-particle
excitation structure as $\mu$. At this stage, we have not 
distinguished between the principal and the auxiliary 
amplitudes. Thus, $\mu$ is a generalised notation 
of the hole-particle excitations, which can be either
principal or auxiliary amplitudes. $D$ is the associated
denominator, usually taken to be the
Hartree-Fock orbital energy difference. As such, in CC 
theory, as shown earlier, there is a bunch of amplitudes
which enslave subsystems composed of auxiliary amplitudes.
In a more generalized case with multiple subsystems, they 
are separately defined by several groups of non-interacting 
auxiliary variables which are enslaved under separate 
interacting groups of the master amplitudes. In the 
simplest case scenario, the iteration
process in CC theory can be viewed as an interplay of 
two interacting subsystems: one containing the principal
amplitudes while
the second one contains only the auxiliary slave amplitudes.
As shown previously via the circular causality relationship,
the master variables (principal amplitudes) determine the
dynamics of the slave variables (auxiliary amplitudes), 
while the updated slave amplitudes pass on their feedback 
to the master amplitudes without affecting the macroscopic 
pattern of the overall dynamics. Here we show how enslaved
amplitudes are governed solely via the master amplitudes. 

As mentioned previously, the amplitudes $t_\mu$ are 
grouped into two classes: the principal amplitudes, 
denoted as $\{t_{L_I}\}$, and the auxiliary amplitudes, 
denoted as \{$t_{S_i}\}$. A general structure of the 
iteration scheme for different principal amplitudes 
(necessarily belonging to the largest subset) may be 
written as:
\begin{equation}
\begin{split}
\Delta t_{L_I}&= \frac{1}{D_{L_I}} (H_{L_I} + {(\contraction{}{H}{}{T_\mu} H T_\mu)}_{L_I} + \frac{1}{2}(\contraction{}{H}{}{T_\mu}
\contraction[2ex]{}{H\;}{T_\nu}{}
H\; T_\mu T_\nu)_{L_I} + . . . )  \\
\end{split}
\label{eq: logical 1}
\end{equation}
As before, the notation $(...)_{L_I}$ means the 
composite inside the parenthesis 
have the hole-particle excitation structure as $L_I$.
Similarly, the equations for determining the enslaved auxiliary
amplitudes, $t_{S_i}$, read as:
\begin{equation}
\begin{split}
\Delta t_{S_i}&= \frac{1}{D_{S_i}} (H_{S_i} + {(\contraction{}{H}{}{T_\mu} H T_\mu)}_{S_i} + \frac{1}{2}(\contraction{}{H}{}{T_\mu}
\contraction[2ex]{}{H\;}{T_\nu}{}
H\; T_\mu T_\nu)_{S_i} + . . . )\\
\end{split}
\label{eq: logical 2}
\end{equation}
Note that till now, the $T_\mu$s appearing in the
second and third terms on the right hand side of the above
equations contain both the principal and auxiliary
amplitudes. One may cast the above two equations 
schematically for the principal and auxiliary amplitudes as:
\begin{equation}
\begin{split}
    D_{L_I}\Delta t_{L_I} &= {(\contraction{}{H^d_{L_I}}{}{T_{L_I}} H^d_{L_I} T_{L_I})}_{L_I} + g_{L_1}(t_{L_1},...,t_{L_{n_L}}, t_{S_1},..., t_{S_{n_S}})\\
   & \forall I= 1, 2, ..., n_L
    \label{eq5}
\end{split}
\end{equation}
and, 
\begin{equation}
\begin{split}
    D_{S_i}\Delta t_{S_1} &= {(\contraction{}{H^d_{S_i}}{}{T_{S_i}} H^d_{S_i} T_{S_i})}_{S_i} + g_{S_1}(t_{L_1},...,t_{L_{n_L}}, t_{S_1},..., t_{S_{n_S}})\\
    & \forall i= 1, 2, ..., n_S
\label{eq6}
\end{split}
\end{equation}
Here we have extracted out the \textit{diagonal} 
contribution from the linear terms, in which the
hole-particle excitation in the cluster operator has the 
same indices as the composite residue. $H^d_\mu$ is thus 
the collection of the Hamiltonian matrix elements associated
with the diagonal term, which \textit{upon contraction} 
to $T_\mu$
produces the composite having same hole-particle 
structure same as $\mu$. Hence to distinguish this term 
from the two body matrix element $H_\mu$ which itself has
hole-particle excitation structure as $\mu$, the
former is depicted by the superscript $d$. Here $g_\mu$ is 
a general nonlinear function that contains the 
constant term, and various linear (except diagonal) and 
all nonlinear terms which arise out of the
Baker-Campbell-Hausdorff expansion. Note that Eqs. \ref{eq5}
and \ref{eq6} have the general structure as:
\begin{equation}
     D_\mu\Delta t_\mu = \gamma_\mu t_\mu + g_\mu(t_{L_1},....,t_{L_{n_L}}, t_{S_1},..., t_{S_{n_S}})
    \label{eq7}
\end{equation}
Here $\gamma_\mu$ in Eq. \ref{eq7} is the generalised
notation of the collection of the Hamiltonian matrix
elements, $H^d_\mu$, which accompanies the diagonal terms 
in the equations for various $t_\mu$'s.
Note that $\gamma_\mu$ has opposite sign to that 
of $D_\mu$ which is evident from their leading order
contributions, and is a measure of the damping constant
associated with the relaxation process of $t_\mu$.
Here we sketchily analyse the leading order contribution 
to the damping constant $\gamma$. 
The leading order contribution to $\gamma_\mu$ in Eqs.
\ref{eq5} and \ref{eq6} is the difference in 
energies involving orbitals of $t_\mu$.
Please see the Appendix for the detailed structure of 
the Hamiltonian matrix element associated with these 
diagonal terms in the orbital basis. Since the inverse of 
the damping constant is proportional to the time scale of
relaxation, one may split the amplitudes into slow and fast 
relaxing modes by analysing the leading order Hamiltonian 
matrix elements.
Note that the principal amplitudes are large in magnitude
and they are labelled by orbitals which are energetically
close. Hence they have smaller damping constants, and 
thus they are characterized by a longer time of relaxation 
to reach their respective fixed points. On the other hand,
the auxiliary amplitudes are labelled by orbitals which are
energetically well separated. The leading term 
contributing to the damping constant, involving the 
Hartree-Fock orbital energy difference, appearing in the
equations for $t_{S_i}, \{i=1,2,...,n_S\}$, is large. Thus 
the auxiliary amplitudes are characterized by larger 
damping constant and a shorter time scale of relaxation.\\
\\
We are now in a position to apply the adiabatic condition
to simplify the working equations. Noting the fact that 
the principal and the auxiliary amplitudes exhibit vastly
different time scale of relaxation, and that 
the variation of the auxiliary amplitudes are suppressed, 
one may invoke an adiabatic approximation by putting
$\Delta t_{S_i}=0, \forall i = 1,2,...,n_S$.
Furthermore, as we argued earlier, we assume that the 
magnitude of the auxiliary amplitudes, $|t_{S_i}|$ is 
much smaller compared to the principal amplitudes:
$|t_{S_i}|<< |t_{L_I}|$. As a consequence, we may put 
$t_{S_i}$ to zero for all values of $i=1,2,...,n_S$ in the 
equations for the auxiliary amplitudes. This allows 
us to write down the equations for the auxiliary 
amplitudes as
\begin{equation}
\begin{split}
    \gamma_{S_i} t_{S_i} + g_{S_i}(t_{L_1},....,t_{L_{n_L}}) = 0 \\
    t_{S_i} = - \gamma_{S_i}^{-1} g_{S_i}(t_{L_1},....,t_{L_{n_L}})
    \label{eq8}
\end{split}
\end{equation}
where $t_{S_1},....,t_{S_{n_S}}$ are put to zero in the
nonlinear terms in Eq. \ref{eq6}. Note that the right hand
side of Eq. \ref{eq8} has only $T_{L_I}$ and hence no 
coupling is allowed among the auxiliary cluster amplitudes
in their equations.

We can now expand the above equations in terms of 
explicit contraction between the Hamiltonian matrix 
elements and the cluster operators which will help us 
in discerning the scaling associated to each term.
The second equation in Eq. \ref{eq8} may be written 
explicitly as:
\begin{equation}
\begin{split}
    t_{S_i} = - {H_{S_i}^{d}}^{-1} (H_{S_i} +\underbracket[0.8pt]{{\Big(\contraction{}{H}{}{T_{L_I}} H T_{L_I})}_{S_i}}_\text{\clap{I~}}  +\underbracket[0.8pt]{\frac{1}{2}(\contraction{}{H}{}{T_{L_I}}
\contraction[2ex]{}{H\;}{T_{L_J}{}}
H \;T_{L_I} T_{L_J})_{S_i}}_\text{\clap{II~}} 
     + . . . \Big)
    \label{eq9}
\end{split}
\end{equation}

Here $\gamma_{S_i}$ has explicitly been written as the 
Hamiltonian matrix element ${H_{S_i}^{d}}$, 
which upon contraction with $T_{S_i}$ produces the 
composite that has the same hole-particle 
excitation as $S_i$. An explicit expression for 
${H_{S_i}^{d}}$ is given in the appendix. We reiterate 
that this term is to be distinguished from the first
term inside the parenthesis in the right-hand side of the 
above equation. Here $H_{S_i}$ means the matrix element 
of Hamiltonian, which has the hole-particle excitation 
structure as $S_i$. Note that in Eq. \ref{eq9}, only the
principal cluster operators $T_{L_I}$ and $T_{L_J}$ appear
in the right hand side, and there is
no term which contains the auxiliary amplitudes. It is 
interesting to note that the auxiliary amplitudes are 
not frozen to predefined values; rather they relax in 
each iteration step via a fixed functional dependence 
on the principal amplitudes as shown in Eq. \ref{eq9}. 

\begin{figure}
\includegraphics[width=\linewidth]{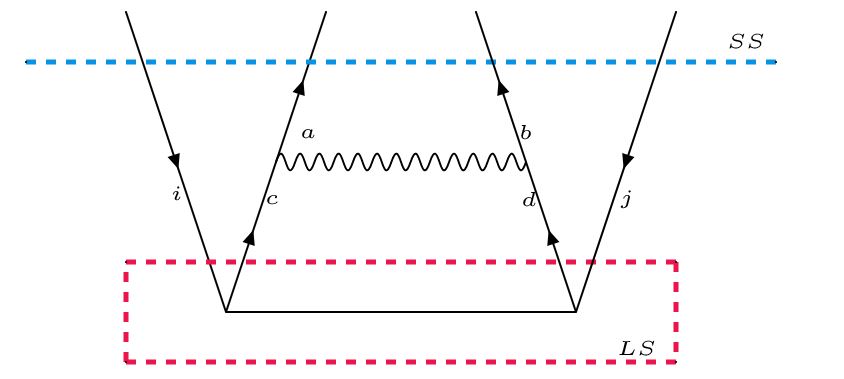}
\caption{The most expensive CC linear diagram appearing from term I, Eq. \ref{eq9}. Note that the set of uncontracted indices, indicated by the blue dashed line, is restricted to those of the auxiliary cluster operators. This is generated via the contraction of the two-electron Hamiltonian matrix element (wiggly line) and the principal cluster operators (solid horizontal line) marked by the red box with dashed line and a subscript $LS$.}
\label{fig:linear}
\end{figure}
\begin{figure}[htp]
\subfloat[A nonlinear diagram appearing from term II, Eq. \ref{eq9}. Note that the set of uncontracted orbital lines, shown by blue dashed line, are fixed to those of the auxiliary cluster operators. The cluster operators shown inside the red boxes with dashed lines are necessarily restricted to the principal operators and hence indicated by subscript $LS$.\label{fig:non-lin-full}]{%
  \includegraphics[clip,width=\linewidth]{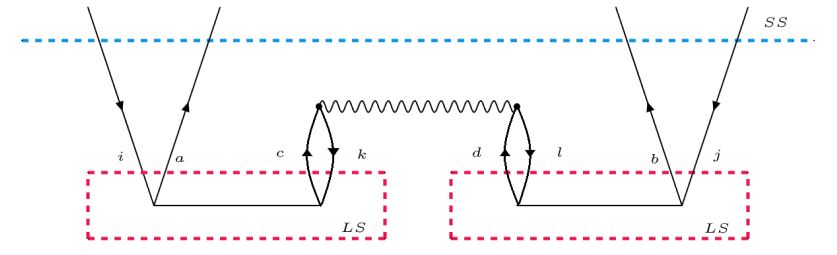}
}%

\subfloat[An intermediate diagram of Fig. \ref{fig:non-lin-full} where the uncontracted lines arising out of the Hamiltonian are restricted to the sets of unique labels of the principal cluster operators and are to be contracted in the next step with a principal cluster operator.]{%
  \includegraphics[clip,width=\linewidth]{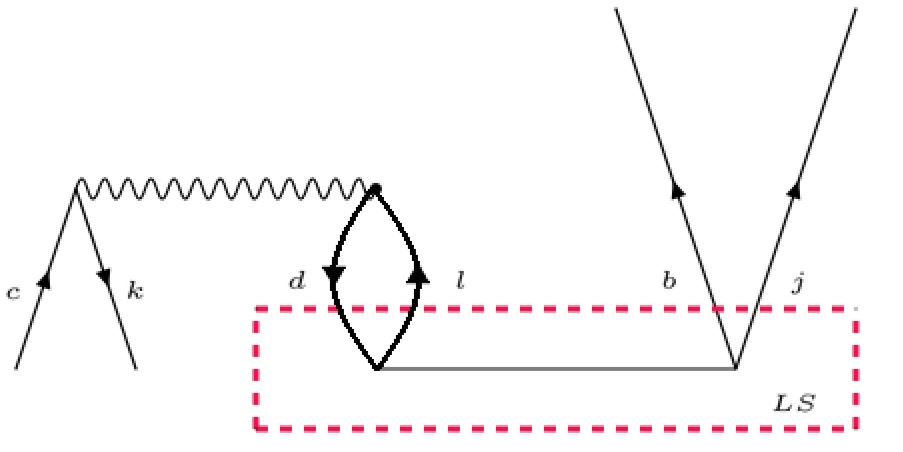}\label{fig:intermediate}%
}
\caption{}

\end{figure}

Let us now consider the scaling associated with the 
construction of the linear and the nonlinear terms in 
Eq. \ref{eq9}, as shown by term I and II, respectively. The 
most expensive diagram in the conventional CCSD arises 
where two particle orbitals contract with the two-body 
cluster operators. The scaling associated with 
the construction of this term is $n_o^2n_v^4$. The 
same term is represented diagrammatically in 
Fig. \ref{fig:linear}. Note that according to term-I 
in Eq. \ref{eq9}, the set of uncontracted indices 
$\{ijab\}$ in Fig. \ref{fig:linear} must correspond 
to the cluster operators belonging to the smaller 
subspace, and this is generated by the contraction 
of the Hamiltonian matrix element with the principal 
cluster operators. Thus 
$S_i = \{ijab\} \in SS$ and $T_{L_I} = T_{ijcd} \in LS$.
This term can be generated at a scaling $n_Ln_v^2$. Note 
that $n_L$ is only a small fraction (usually 10\%-20\%, 
\textit{vide infra}) of the total number of cluster
operators. This leads to an order of magnitude reduction 
in scaling in constructing the most expensive CC diagram. 

We now turn our attention to the construction of the 
nonlinear terms as indicated by II in Eq. \ref{eq9}. Let
us consider the diagram as shown by Fig. \ref{fig:non-lin-full}. As it was in the
case for the linear term, here also the set of uncontracted 
indices corresponds to the auxiliary cluster operators. 
Thus, $S_i = \{ijab\} \in SS$. This term is generated via 
the contraction of the Hamiltonian with two principal
cluster operators: $T_{L_I} = T_{ikac} \in LS$, and 
$T_{L_J} = T_{jlbd} \in LS$. Like the conventional 
CC theory, this too can be constructed via an optimal
intermediate shown in Fig. \ref{fig:intermediate}. 
In order to generate this intermediate,
one may restrict the pair of orbitals $k$ and 
$c$ to unique pair labels that appear in one of the 
vertices of the principal 
operators (in the LS). This is a small fraction of terms
compared to keeping $n_on_v$ number of terms in an 
unrestricted manner. Thus the construction of the 
intermediate takes $n_Ln_u$ matrix operations.
Here $n_u=dim(kc \in LS)_{unique}$, which is 
smaller than $n_on_v$. Thus the overall computational
scaling to construct the intermediate, $n_Ln_u$, is 
much smaller than the conventional $n_o^3n_v^3$ as
required in the conventional CC. The construction
of the final diagram via this intermediate and $T_{ikac}$
takes the same computational scaling. Thus the 
overall scaling for the generation of the auxiliary
amplitudes are reduced by at least an order of 
magnitude compared to the conventional CCSD. We now focus 
on the feedback coupling to determine the principal cluster
amplitudes as shown in step II of the circular causality
diagram.

\subsection{The Feedback coupling: Determination of the principal driver amplitudes:}
As we had shown in the earlier section, due to the 
difference in the time scale, the auxiliary amplitudes can
be decoupled from the principal amplitudes by demanding 
that the variation of the former is suppressed during the
iteration process. This, coupled with the elimination of the
auxiliary amplitudes from their equations due to their 
smallness in magnitude, allows us to write the auxiliary 
amplitudes in terms of the functionals of the principal
amplitudes alone. We now focus on the determination 
of the principal amplitudes. Noting the fact that each 
auxiliary amplitudes are functionals of the principal
amplitudes, one may now rewrite Eq. \ref{eq5} as:
\begin{equation}
    \begin{split}
    D_{L_I}\Delta t_{L_I} &= {(\contraction{}{H^d_{L_I}}{}{T_{L_I}} H^d_{L_I} T_{L_I})}_{L_I} + g_{L_1}\Big(t_{L_1},...,t_{L_{n_L}},\\& t_{S_1}(t_{L_1},...,t_{L_{n_L}}),..., t_{S_{n_S}}(t_{L_1},...,t_{L_{n_L}})\Big)\\
    & \forall I= 1, 2, ..., n_L
    \end{split}
    \label{eq10}
\end{equation}

Here we have explicitly shown that the auxiliary 
amplitudes appearing in various $g_{L_I}$ are functions 
of the principal amplitudes. Absorbing the diagonal term,
one may recast Eq. \ref{eq: logical 1} as:
\begin{multline}
\Delta t_{L_I}= \frac{1}{D_{L_I}} \Big(\underbracket[0.8pt]{H_{L_I}}_\text{\clap{I~}} +
\underbracket[0.8pt]{{(\contraction{}{H}{}{T_{L_J}} H T_{L_J})}_{L_I}}_\text{\clap{II~}} +
\underbracket[0.8pt]{{(\contraction{}{H}{}{T_{S_i}} H T_{S_i})}_{L_I}}_\text{\clap{III~}}+ \\  \frac{1}{2}\big(\underbracket[0.8pt]{(\contraction{}{H}{}{T_{L_J}} \contraction[2ex]{}{H\;}{T_{L_K}}{} H\; T_{L_J} T_{L_K})_{L_I}}_\text{\clap{IV~}}
+\underbracket[0.8pt]{(\contraction{}{H}{}{T_{L_J}} \contraction[2ex]{}{H\;}{T_{S_i}}{} H\; T_{L_J} T_{S_i})_{L_I}}_\text{\clap{V~}}+\underbracket[0.8pt]{(\contraction{}{H}{}{T_{S_i}} \contraction[2ex]{}{H\;}{T_{S_j}}{} H\; T_{S_i} T_{S_j})_{L_I}}_\text{\clap{VI~}}\big)+
. . . \Big) 
\label{eq:feedback-coupling}
\end{multline} with 
\begin{equation}
    T_{S_i}=T_{S_i}[\{T_L\}]
\end{equation}
Here, we have explicitly shown the different combinations 
of the linear and the
nonlinear terms that appear with various powers of the 
principal and auxiliary cluster amplitudes. As we argued 
above, the auxiliary amplitudes are written
explicitly as functional of the principal amplitudes, and 
they are determined via Eq. \ref{eq9}. Depending
on which set of terms are retained in the equations of the 
principal amplitudes in Eq. \ref{eq:feedback-coupling}, 
we have devised two schemes having different levels of 
sophistication, as discussed below.

In scheme-I, we have kept all the terms (I-VI) in Eq. 
\ref{eq:feedback-coupling}, and we will refer this scheme as
AD-CC (Scheme-I). This thus introduces complete 
feedback of the auxiliary amplitudes to the equations for
the principal amplitudes, and hence the principal 
amplitudes are determined \textit{exactly}. On the other 
hand, since the auxiliary amplitudes are much smaller in
magnitude compared to the principal amplitudes, one may 
selectively include only those terms in Eq. 
\ref{eq:feedback-coupling} where the auxiliary amplitudes
are kept only up to the linear power. This implies that 
all the nonlinear terms
which include auxiliary amplitudes may safely be neglected.
Thus in AD-CC (Scheme-II), we have retained terms (I-IV),
and have neglected terms (V-VI) in Eq. 
\ref{eq:feedback-coupling} for determining the principal 
cluster amplitudes. We will show in the next section that
although this may seem to be a rather drastic 
approximation, the accuracy
of the results does not deteriorate much compared to 
scheme-I. In fact, 
AD-CC (Scheme-II) may be chosen as an excellent cost 
effective way to solve CC theory without scarifying the 
accuracy significantly. It will also be pretty evident that 
the results are systematically improvable by going from 
scheme-II to scheme-I, since the latter has a more complete 
expression for the principal amplitudes. One may also design 
a mid-way scheme where all the terms (I-V) are included and
only term VI is neglected in Eq. \ref{eq:feedback-coupling}.
However, in the present work, we have not considered this 
alternative. Below, we discuss the computational scaling 
for the determination of the principal amplitudes via 
both the schemes.

\begin{figure}
    \centering
    \includegraphics[width=\linewidth]{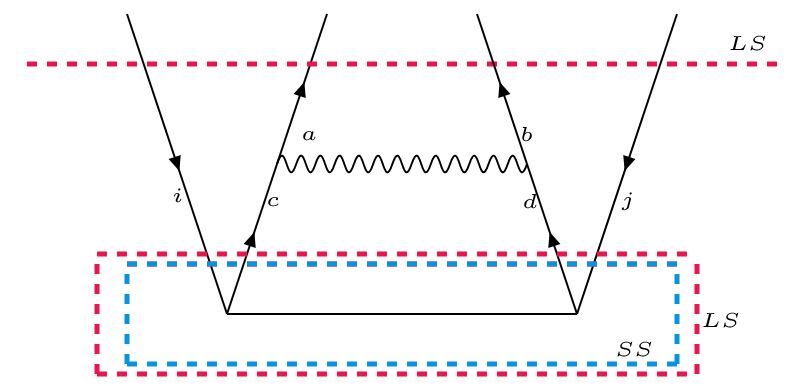}
    \caption{A linear diagram arising from terms II and III, Eq. \ref{eq:feedback-coupling}. Note that the cluster operator can be either a principal operator (red box, coming from term II, Eq. \ref{eq:feedback-coupling} or an auxilliary operator (blue box, coming from term III, Eq. \ref{eq:feedback-coupling}. The set of uncontracted indices are necessarily restricted to the elements of LS}
    \label{fig:Lin-T-l}
\end{figure}

Note that in both schemes-I and scheme-II, both the 
linear terms, 
II and III, are included in Eq. \ref{eq:feedback-coupling}.
This implies that both the principal and the auxiliary 
cluster operators contract to the Hamiltonian to
generate a composite structure where the uncontracted 
indices may take only those values which together
characterize a principal cluster operator. This is 
diagrammatically shown via the most expensive CC diagram
in Fig. \ref{fig:Lin-T-l}. Note that in Fig. \ref{fig:Lin-T-l}, unlike Fig. 
\ref{fig:linear}, the set of uncontracted
indices $\{ijab\}$ necessarily characterize a principal
cluster operator, whereas the horizontal solid line can be 
both principal and auxiliary cluster operators. Since
the uncontracted indices are restricted to the LS
elements, one trivially can construct this term at a
scaling of $n_L n_v^2$. Although the diagram shown in 
Fig. \ref{fig:Lin-T-l} is superficially same as that of 
Fig. \ref{fig:linear}, note that they have different
interpretations in two different contexts.
\begin{figure}[htp]

\subfloat[A nonlinear diagram appearing from terms IV, V, VI, Eq. \ref{eq:feedback-coupling}. Note that the set of uncontracted orbital lines, shown by red dashed line, are fixed to those of the auxiliary cluster operators. The cluster operators are shown inside dashed boxes, and they can be the principal amplitudes (the red box) as well as auxiliary amplitudes (blue box) for scheme-I. For scheme-II, the cluster operators can only be the principal operator and the auxiliary operators (the blue box) do not exist.]{%
  \includegraphics[clip,width=\linewidth]{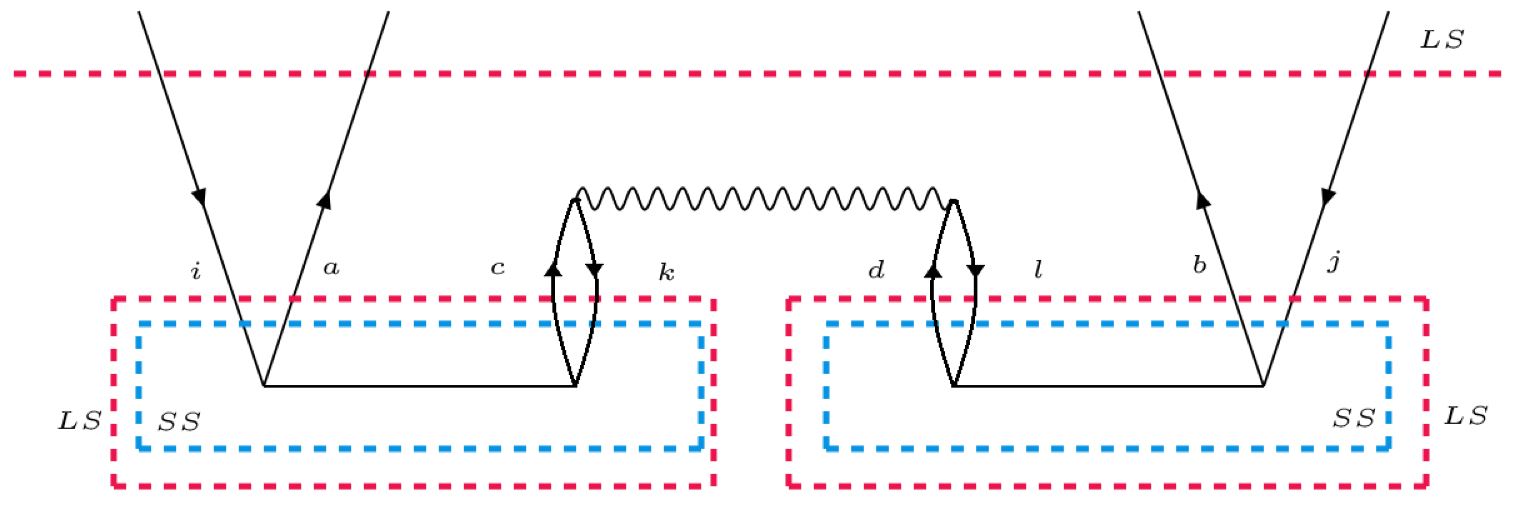}%
  \label{fig:non-lin-T-l}
}

\subfloat[An intermediate diagram of Fig. \ref{fig:non-lin-T-l} where the uncontracted lines arising out of the cluster operator (for both scheme-I and II) are restricted to the sets of unique labels of the principal cluster operators. For scheme-II, the pair of lines coming out of the Hamiltonian are restricted to unique labels of principal cluster operators, whereas they can take any hole and particle labels for scheme-I.]{%
  \includegraphics[clip,width=\linewidth]{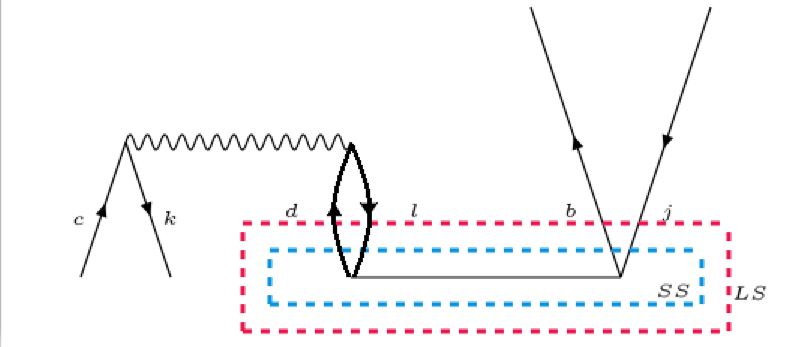}%
 \label{fig:int-tl}
}
\caption{}

\end{figure}

\begin{figure}[!b]
     \centering
     \includegraphics[width=\linewidth]{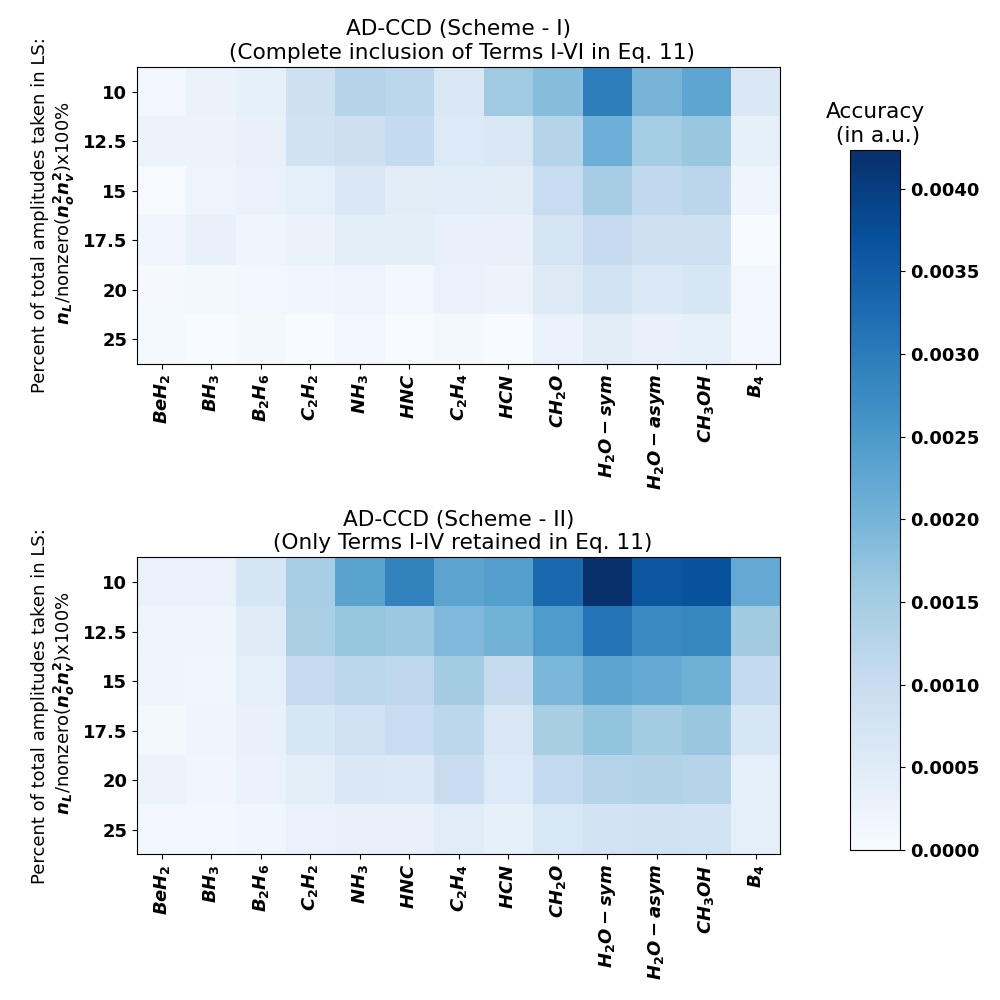}
     \caption{Difference in energy of the AD-CCD schemes (from the canonical CCD) for different molecular systems with varied electronic complexity in cc-pVDZ basis as a function of the LS dimension, $n_L$. The lighter shades signify higher accuracy. Note that scheme-I shows sub milliHartree($mE_H$) accuracy with $n_L$ taken to be 10\% of the total nonzero cluster amplitudes. The results systematically improve with increase in LS dimension. Scheme-II, owing to the incomplete coupling of the auxiliary amplitudes to the equations of the principal amplitudes, require slightly larger LS dimension to achieve similar accuracy.}
     \label{fig:heatmap}
\end{figure}
We now turn our attention to analyse the scaling of the 
nonlinear diagrams in Eq. \ref{eq:feedback-coupling}. One
may again construct diagrams of similar topology like those 
appearing in Fig. \ref{fig:non-lin-full}, and interpret it 
in the context of Eq. \ref{eq:feedback-coupling}. Here we 
analyse a diagram as shown in Fig. \ref{fig:non-lin-T-l},
having similar topology to that of Fig. \ref{fig:non-lin-full}.
Note that like the linear term,
the set of uncontracted index $\{ijab\}$ can take only
those elements which characterize a principal cluster
amplitude. However, in AD-CC (scheme-I), 
the cluster operators $T_{ikac}$ and $T_{jlbd}$
can be both principal and auxiliary operators as evident from
Eq. \ref{eq:feedback-coupling}. Since the set of uncontracted
indices belong to LS, in order to construct
the intermediate as shown in Fig \ref{fig:int-tl}, one may restrict
the pair of orbitals $j$ and $b$ to unique pair labels 
which appear in one of the vertices of the principal 
cluster operators. Note that the orbitals $k, l$ and $c, d$
can take up any value from the entire set of occupied 
and unoccupied orbitals respectively. Thus to construct the
intermediate, one needs to perform 
$n_o^2 n_v^2 n_u$ matrix operations, which is less 
than $n_o^3 n_v^3$ matrix operations needed to construct 
the same intermediate in the conventional CC theory. The next
step to construct the entire diagram takes $n_L n_o n_v$ 
computational scaling.

On the other hand, in AD-CC (scheme-II), the cluster
operators $T_{ikac}$ and $T_{jlbd}$ can only be the 
principal cluster operators. That implies that the
intermediate, as shown in Fig \ref{fig:int-tl}, can be
constructed with a computational
scaling of $n_L n_u$. Note that
this is one to two order of magnitude reduction compared to
the conventional $n_o^3 n_v^3$ scaling necessary to construct
the diagram. In the next section, we will present a few
prototypical numerical results to demonstrate the efficacy
of both the schemes. We will also demonstrate the relative 
scaling of our schemes for the construction of the linear 
and nonlinear terms that appear in the forward mapping 
(step-I, Fig. \ref{fig:circ-caus}) and backward mapping
(step-II, Fig. \ref{fig:circ-caus}) compared to the 
conventional CCD scheme.
\section{Results and Discussion}
In this section, we demonstrate the efficacy of the AD-CC
(scheme-I/II) with a number of molecular systems. As the 
proof of the concept, we will restrict our applications to
adiabatically decoupled CCD (AD-CCD) theory, although the 
generalization to include singles or triples and higher
excitations is fairly straightforward. All the calculations
were done using our in-house CC codes, and the convergence 
threshold for all the calculations was set to be 
$1\times 10^{-6}$. No DIIS routine was used to accelerate 
the calculations although doing so is fairly straightforward. 

\onecolumngrid
\begin{figure*}
\includegraphics[width=\textwidth]{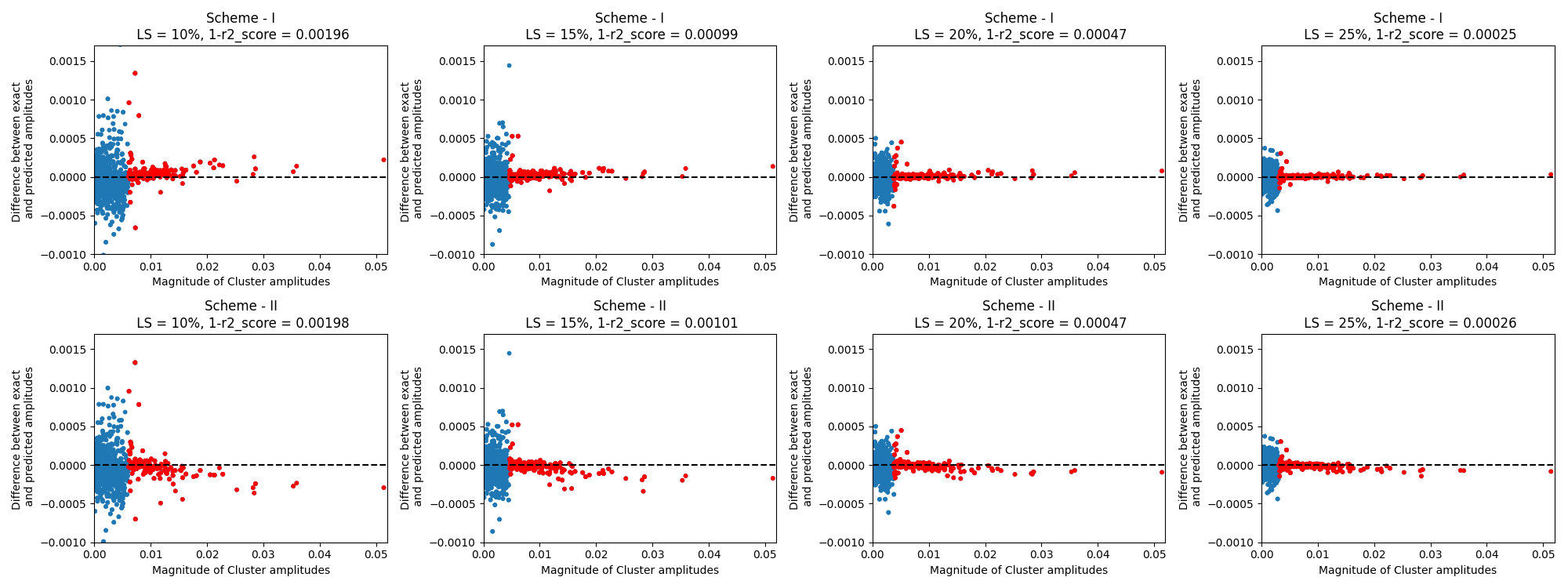}
\caption{The difference between the predicted amplitudes and the exact amplitudes (obtained from Eqs. \ref{eq9} and \ref{eq:feedback-coupling}) are plotted for Water. The red dots denote the cluster amplitudes that were taken in the largest subset (Eq. \ref{eq:feedback-coupling}), and the blue are the remaining smaller amplitudes
    (obtained via Eq. \ref{eq9}). The LS elements are predicted much more accurately as compared to the SS. Both show systematic improvement as the LS dimension is increased.}
    \label{fig:scatter}
\end{figure*}
\begin{table*}
\centering
\begin{tabular}{|p{6.5cm}|M{0.8cm}|M{0.8cm}|M{0.8cm}|M{0.8cm}|M{0.8cm}|M{0.9cm}|M{0.8cm}|M{0.9cm}|M{0.9cm}|M{0.9cm}|M{0.8cm}|M{0.9cm}|}
\hline 
\diagbox[width=\linewidth]{Relative\\ Scaling}{Molecule} & \multicolumn{4}{C{3.2cm}|}{Ammonia  ($n_o=5,n_v=24$)  Asymmetric } &\multicolumn{4}{C{3.3cm}|}{Methanol ($n_o=9,n_v=39$) (slightly symmetric)} & \multicolumn{4}{C{3.4cm}|}{Ethene ($n_o=8,n_v=40$) (highly symmetric)}\\ 
\hline 
LS Size (\% of Non zero elements) & 10 & 15 & 20 & 25 & 10 & 15 & 20 & 25 & 10 & 15 & 20 & 25 \\ \hline
$n_u = dim(jb\in LS)_{unique}$& 95 & 96 & 98 & 99& 266 & 274 & 279 & 282& 176 & 209 & 225 & 234\\ \hline 
Linear term construction 
(Eq. \ref{eq9}, term-I, Fig. \ref{fig:linear}) 
AD-CCD:CCD = $n_Ln_v^2/ n_o^2n_v^4$ 
& 0.1 & 0.15 & 0.2 & 0.25 & 0.05 & 0.075 & 0.1 & 0.125 & 0.01 & 0.015 & 0.02 & 0.025 \\  \hline 
Linear term construction 
(Eq. \ref{eq:feedback-coupling}, term-II +III, Fig. \ref{fig:Lin-T-l}) 
AD-CCD:CCD = $n_Ln_v^2/ n_o^2n_v^4$ 
& 0.1 & 0.15 & 0.2 & 0.25 & 0.05 & 0.075 & 0.1 & 0.125 & 0.01 & 0.015 & 0.02 & 0.025 \\  \hline
Intermediate term construction 
(for term-II in Eq. \ref{eq9}; Fig. \ref{fig:intermediate}) 
AD-CCD:CCD = $n_L n_u /n_o^3n_v^3$ 
& 0.08 & 0.12 & 0.163 & 0.206 & 0.037 & 0.0573& 0.077 & 0.0975 & 0.0055 & 0.01 & 0.014 & 0.0183 \\ \hline
Intermediate term construction 
(for terms-IV+V+VI in Eq. \ref{eq:feedback-coupling}; Fig. \ref{fig:int-tl}) 
AD-CCD (Scheme-I):CCD: $n_o^2n_v^2n_u /n_o^3n_v^3$ 
& 0.79 & 0.8 & 0.816 & 0.825 & 0.74 & 0.76 & 0.773 & 0.78 & 0.55&0.653&0.70&0.73 \\ \hline
Intermediate term construction 
(for term-IV in Eq. \ref{eq:feedback-coupling}; Fig. \ref{fig:int-tl})
AD-CCD (Scheme-II):CCD = $n_L n_u /n_o^3n_v^3$ 
& 0.08 & 0.12 & 0.163 & 0.206 & 0.037 & 0.057& 0.077 & 0.0975 & 0.0055 & 0.01 & 0.014 & 0.018 \\ \hline 
\end{tabular} 
\caption{Relative computational scaling for AD-CCD scheme-I and II for construction of linear diagrams and intermediates to nonlinear diagrams in step-I and step-II of the circular causality loop. The fractions are the ratio of the computational scaling required by AD-CCD schemes and conventional CCD to construct diagrams or intermediates of same topological structures for three molecules in their equilibrium and away from equilibrium geometries in cc-pVDZ basis.}
\label{tab1}
\end{table*}
\twocolumngrid
In Fig. \ref{fig:heatmap}, we have demonstrated the accuracy
of our models for a few molecular systems with varied 
degree of complexity and size. In  the vertical axis, we 
plot the dimension of the largest subset, which is chosen
to vary from 10\% to 25\% of the total nonzero cluster 
amplitudes. We reiterate once again that the choice of the
LS is somewhat not very elegant, although for all the 
practical purposes, this is fairly accurate. In Fig. \ref{fig:heatmap}, the colour 
coding of the individual boxes
represent the difference of AD-CCD schemes from the canonical
CCD calculations. The darker shades imply a larger deviation,
while the lighter shades imply a smaller difference. 

The upper panel of Fig. \ref{fig:heatmap} shows the accuracy 
of a few molecules in their equilibrium and away from 
equilibrium geometries in cc-pVDZ basis in AD-CCD scheme-I, 
while the lower panel shows the accuracy for scheme-II.
Clearly, for all the molecules under consideration, the 
AD-CCD (scheme-I) shows sub milliHartree ($mE_H$) accuracy
compared to the conventional CCD with the LS
chosen to be the largest 
15\% of the total cluster amplitudes. The method, as 
expected, shows a systematic improvement in accuracy as 
one includes more elements in the LS. With the LS dimension
of 20\% of the total cluster amplitudes and above, the method,
on an average, shows accuracy of the order of
0.1 $mE_H$. Since the AC-CCD (scheme-II) includes an 
incomplete coupling of the auxiliary amplitudes in the
equations for the principal amplitudes, one expects somewhat
less accuracy, particularly with smaller size of the LS. This
is clearly evident that one needs to increase the LS 
dimension to 20\% to achieve a sub-$mE_H$ accuracy in
AD-CCD (scheme-II). Nonetheless, scheme-II shows, on an 
average, deviation of 0.5 $mE_H$ with the LS taken to be the
20\% of the total number of nonzero cluster amplitudes. On
the other hand, AD-CCD (scheme-I) requires 15\% of the 
total nonzero cluster amplitudes to achieve a similar 
accuracy.

In order to demonstrate the exactness of the cluster 
amplitudes, in Fig. \ref{fig:scatter}, we have plotted 
the difference of the exact CCD amplitudes and those
predicted via Eqs. \ref{eq9} and \ref{eq:feedback-coupling}
for water molecule at equilibrium in cc-pVDZ basis.
The red dots are the principal amplitudes obtained via 
Eq. \ref{eq:feedback-coupling}, while the blue dots 
signify the auxiliary cluster amplitudes obtained 
from Eq. \ref{eq9}. Clearly, the smaller amplitudes get
more and more accurate as one increases the LS dimension.
This in turn results in accurate prediction of the principal 
amplitudes as well, since the auxiliary amplitudes provide 
the feedback coupling to the principal amplitudes via Eq.
\ref{eq:feedback-coupling}. It is to be noted that with
increase in the LS dimension, there is a systematic 
improvement of the accuracy of the predicted cluster 
amplitudes with the overall R2 score of 0.99804 at 10\% 
to 0.99975 at 25\% with AD-CCD (scheme-I). The 
approximated AD-CCD (scheme-II) also shows quite a
similar trend (R2 score=0.99974 at 25\%), although
it is marginally less accurate with smaller 
dimension of the LS (R2 score=0.99802 at
10\%). This is quite expected as scheme-II uses a 
drastically approximated equation to determine the principal
amplitudes where all the nonlinear terms containing the 
auxiliary amplitudes are neglected. However, with increase 
in the LS dimension, both the schemes tend to produce 
similar results as shown over a number of molecular 
applications in Fig. \ref{fig:heatmap}. We reiterate 
that the remarkable accuracy of AD-CCD, scheme-I in
particular, has been achieved with an order of magnitude 
reduction in computational scaling which will be 
demonstrated below.

Lastly, we turn our attention to demonstrate the reduction
in computational scaling of both the schemes over the 
conventional CCD. We would be presenting with three 
different molecular systems having varied degrees of 
complexity. In Table \ref{tab1}, we have presented the 
ratio of the computational scaling required for the forward
and backward mappings (step-I and step-II, of Fig.
\ref{fig:circ-caus}) for both the 
schemes to that of the conventional CCD in order to 
construct the most expensive linear diagram (and the
intermediates to the nonlinear diagrams). The rate 
determining step to the conventional CCD calculations 
is the construction of the linear diagram where two 
particle orbitals contract. The scaling associated 
to this diagram is $n_o^2n_v^4$. Construction of the 
diagram having same topological structure in both
the forward mapping and the feedback coupling scales as
$n_L n_v^2$. For the highly distorted molecule without
any symmetry, the dimension of the LS, $n_L$, is at max
taken to be 10-25\% of the total (nonzero) cluster 
amplitudes for sub $mE_H$ accuracy. This leads to 
at least 75\%-90\% reduction in computational scaling 
for the construction of the most expensive CCD diagram 
which has two particle contraction. We now analyse 
the computational cost associated with the construction 
of the intermediates, which in conventional CCD scales as
$n_o^3n_v^3$. For AD-CCD (scheme-I), as we have shown in 
Fig. \ref{fig:non-lin-T-l}, we have a 
lower scaling of $n_u n_o^2 n_v^2$, where $n_u$ is the
number of unique excitation vertex that appears in the 
two-body principal cluster operators in the LS. The 
variation of $n_u$ is shown in Table \ref{tab1} for
different choices of the LS dimension and for different 
molecules. Clearly, $n_u$ varies sub-linearly with the 
increase in LS dimension, $n_L$. Usually, $n_u$ is about
70-80\% of $n_on_v$. Thus, the construction of the 
intermediate in Fig. \ref{fig:non-lin-T-l} takes 
computational scaling of about 
$(0.70-0.80) \times n_o^3 n_v^3$. For AD-CCD (scheme-II), 
on the other hand, allows only the principal amplitudes 
in the nonlinear term of the feedback coupling (step-II of 
Fig. \ref{fig:circ-caus}). Thus, it further 
reduces the computational scaling to $n_Ln_u$, which is only
about 8-20\% of $n_o^3 n_v^3$. However, as evident from 
Fig. \ref{fig:heatmap}, this does not deteriorate the 
accuracy of AD-CCD (scheme-II), particularly with a
conservative choice of the LS. We should 
note that for AD-CCD (scheme-I), in a small 
basis set, the scaling required for the construction 
of the intermediate for the 
nonlinear term is often comparable to the computational
scaling required to construct the linear diagram with
two particle contraction. On the other hand, 
in large basis sets, 
the construction of the linear diagram is the rate
determining step. However, we reiterate that in our 
AD-CCD schemes, the scaling for the construction of this 
is drastically reduced by 70-80\%.

\section{Conclusion and Future Outlook}
In this article, we had developed a novel variant of 
an approximated CC theory where the solution of two sets 
of the cluster amplitudes are decoupled. The method
relies upon an adiabatic approximation where the fast 
relaxing stable amplitudes are decoupled and expressed as
fixed functionals of the slow moving unstable amplitudes.
The principal amplitudes, which are the unstable modes 
of the iteration dynamics, are determined accurately 
via exact CC equations that include the coupling of 
all unstable amplitudes and a feedback coupling of 
the stable auxiliary amplitudes. Two approximated 
schemes have been proposed which differ in the level of the
inclusion of the nonlinear terms containing the auxiliary
amplitudes. As a proof of the concept, we have included
only the double excitations in our applications. 
Both the resulting schemes of AD-CCD have been applied to a 
number of molecules in their equilibrium and away from
equilibrium geometries, and they show remarkable accuracy
compared to the conventional
CCD. While the typical accuracy of the methods depend on
the dimension of the chosen LS, a conservative choice of 
the LS leads to the results which is accurate up to the 
order of one hundredth of a $mE_H$. This is achieved
with a computational scaling of the order of 10\%-25\% of 
the typical $n_o^2n_v^4$ scaling of CCD, at worst.

While the current choice of the LS is fairly general and 
works for all practical purposes, this is not the optimal
choice by any means. One may rely upon Lyapunov stability 
analysis or information theoretic techniques to choose 
the optimal LS and reduce the computational scaling even
further. Moreover, a generalization to the current work 
to include triples and higher excitations is 
straightforward, and will be subject to one of our 
forthcoming publications. 

\section{Acknowledgments}
The authors thank Mr. Anish Chakraborty, IIT 
Bombay, for many stimulating discussions about the
structure of the program. RM acknowledges the 
financial support from Industrial Research and
Consultancy Centre, IIT Bombay, and 
Science and Engineering Research Board, Government
of India, for financial support.

\section*{Data Availability}
The data is available upon reasonable request to the corresponding author.

\section{APPENDIX:}

\begin{multline}
    H_{S_i}^d = (1 + P(i,j)P(a,b)) \Big(-f_{ii}+f_{aa}+\frac{1}{2}v_{ab}^{ab}+2v_{ia}^{ai}\\-(1+\delta_{ij}\delta_{ab}) v_{ia}^{ia}+ \frac{1}{2}v_{ij}^{ij}- v_{ib}^{ib} \Big)\\
    S_i = \{ijab\} \in SS
\end{multline}    


\begin{thebibliography}{17}%
\makeatletter
\providecommand \@ifxundefined [1]{%
 \@ifx{#1\undefined}
}%
\providecommand \@ifnum [1]{%
 \ifnum #1\expandafter \@firstoftwo
 \else \expandafter \@secondoftwo
 \fi
}%
\providecommand \@ifx [1]{%
 \ifx #1\expandafter \@firstoftwo
 \else \expandafter \@secondoftwo
 \fi
}%
\providecommand \natexlab [1]{#1}%
\providecommand \enquote  [1]{``#1''}%
\providecommand \bibnamefont  [1]{#1}%
\providecommand \bibfnamefont [1]{#1}%
\providecommand \citenamefont [1]{#1}%
\providecommand \href@noop [0]{\@secondoftwo}%
\providecommand \href [0]{\begingroup \@sanitize@url \@href}%
\providecommand \@href[1]{\@@startlink{#1}\@@href}%
\providecommand \@@href[1]{\endgroup#1\@@endlink}%
\providecommand \@sanitize@url [0]{\catcode `\\12\catcode `\$12\catcode
  `\&12\catcode `\#12\catcode `\^12\catcode `\_12\catcode `\%12\relax}%
\providecommand \@@startlink[1]{}%
\providecommand \@@endlink[0]{}%
\providecommand \url  [0]{\begingroup\@sanitize@url \@url }%
\providecommand \@url [1]{\endgroup\@href {#1}{\urlprefix }}%
\providecommand \urlprefix  [0]{URL }%
\providecommand \Eprint [0]{\href }%
\providecommand \doibase [0]{http://dx.doi.org/}%
\providecommand \selectlanguage [0]{\@gobble}%
\providecommand \bibinfo  [0]{\@secondoftwo}%
\providecommand \bibfield  [0]{\@secondoftwo}%
\providecommand \translation [1]{[#1]}%
\providecommand \BibitemOpen [0]{}%
\providecommand \bibitemStop [0]{}%
\providecommand \bibitemNoStop [0]{.\EOS\space}%
\providecommand \EOS [0]{\spacefactor3000\relax}%
\providecommand \BibitemShut  [1]{\csname bibitem#1\endcsname}%
\let\auto@bib@innerbib\@empty
\bibitem [{\citenamefont {\u{C}\'{i}\u{z}ek}(1966)}]{cc3}%
  \BibitemOpen
  \bibfield  {author} {\bibinfo {author} {\bibfnamefont {J.}~\bibnamefont
  {\u{C}\'{i}\u{z}ek}},\ }\href@noop {} {\bibfield  {journal} {\bibinfo
  {journal} {J. Chem. Phys.}\ }\textbf {\bibinfo {volume} {45}},\ \bibinfo
  {pages} {4256} (\bibinfo {year} {1966})}\BibitemShut {NoStop}%
\bibitem [{\citenamefont {\u{C}\'{i}\u{z}ek}(1969)}]{cc4}%
  \BibitemOpen
  \bibfield  {author} {\bibinfo {author} {\bibfnamefont {J.}~\bibnamefont
  {\u{C}\'{i}\u{z}ek}},\ }\href@noop {} {\bibfield  {journal} {\bibinfo
  {journal} {Adv. Chem. Phys.}\ }\textbf {\bibinfo {volume} {14}},\ \bibinfo
  {pages} {35} (\bibinfo {year} {1969})}\BibitemShut {NoStop}%
\bibitem [{\citenamefont {{\v{C}}{\'\i}{\v{z}}ek}\ and\ \citenamefont
  {Paldus}(1971)}]{cc5}%
  \BibitemOpen
  \bibfield  {author} {\bibinfo {author} {\bibfnamefont {J.}~\bibnamefont
  {{\v{C}}{\'\i}{\v{z}}ek}}\ and\ \bibinfo {author} {\bibfnamefont
  {J.}~\bibnamefont {Paldus}},\ }\href@noop {} {\bibfield  {journal} {\bibinfo
  {journal} {Int. J. Quantum Chem.}\ }\textbf {\bibinfo {volume} {5}},\
  \bibinfo {pages} {359} (\bibinfo {year} {1971})}\BibitemShut {NoStop}%
\bibitem [{\citenamefont {Bartlett}\ and\ \citenamefont
  {Musia{\l}}(2007)}]{bartlett2007coupled}%
  \BibitemOpen
  \bibfield  {author} {\bibinfo {author} {\bibfnamefont {R.~J.}\ \bibnamefont
  {Bartlett}}\ and\ \bibinfo {author} {\bibfnamefont {M.}~\bibnamefont
  {Musia{\l}}},\ }\href@noop {} {\bibfield  {journal} {\bibinfo  {journal}
  {Reviews of Modern Physics}\ }\textbf {\bibinfo {volume} {79}},\ \bibinfo
  {pages} {291} (\bibinfo {year} {2007})}\BibitemShut {NoStop}%
\bibitem [{\citenamefont {Szak{\'a}cs}\ and\ \citenamefont
  {Surj{\'a}n}(2008{\natexlab{a}})}]{szakacs2008iterative}%
  \BibitemOpen
  \bibfield  {author} {\bibinfo {author} {\bibfnamefont {P.}~\bibnamefont
  {Szak{\'a}cs}}\ and\ \bibinfo {author} {\bibfnamefont {P.~R.}\ \bibnamefont
  {Surj{\'a}n}},\ }\href@noop {} {\ \textbf {\bibinfo {volume} {43}},\ \bibinfo
  {pages} {314} (\bibinfo {year} {2008}{\natexlab{a}})}\BibitemShut {NoStop}%
\bibitem [{\citenamefont {Szak{\'a}cs}\ and\ \citenamefont
  {Surj{\'a}n}(2008{\natexlab{b}})}]{szakacs2008stability}%
  \BibitemOpen
  \bibfield  {author} {\bibinfo {author} {\bibfnamefont {P.}~\bibnamefont
  {Szak{\'a}cs}}\ and\ \bibinfo {author} {\bibfnamefont {P.~R.}\ \bibnamefont
  {Surj{\'a}n}},\ }\href@noop {} {\bibfield  {journal} {\bibinfo  {journal}
  {Int. J. Quantum Chem.}\ }\textbf {\bibinfo {volume} {108}},\ \bibinfo
  {pages} {2043} (\bibinfo {year} {2008}{\natexlab{b}})}\BibitemShut {NoStop}%
\bibitem [{\citenamefont {Maitra}\ and\ \citenamefont
  {Nakajima}(2017)}]{maitra_correlation_2017}%
  \BibitemOpen
  \bibfield  {author} {\bibinfo {author} {\bibfnamefont {R.}~\bibnamefont
  {Maitra}}\ and\ \bibinfo {author} {\bibfnamefont {T.}~\bibnamefont
  {Nakajima}},\ }\href {\doibase 10.1063/1.5000571} {\bibfield  {journal}
  {\bibinfo  {journal} {J. Chem. Phys.}\ }\textbf {\bibinfo {volume} {147}},\
  \bibinfo {pages} {204108} (\bibinfo {year} {2017})}\BibitemShut {NoStop}%
\bibitem [{\citenamefont {Maitra}, \citenamefont {Akinaga},\ and\ \citenamefont
  {Nakajima}(2017)}]{maitra_coupled_2017}%
  \BibitemOpen
  \bibfield  {author} {\bibinfo {author} {\bibfnamefont {R.}~\bibnamefont
  {Maitra}}, \bibinfo {author} {\bibfnamefont {Y.}~\bibnamefont {Akinaga}}, \
  and\ \bibinfo {author} {\bibfnamefont {T.}~\bibnamefont {Nakajima}},\ }\href
  {\doibase 10.1063/1.4985916} {\bibfield  {journal} {\bibinfo  {journal} {J.
  Chem. Phys.}\ }\textbf {\bibinfo {volume} {147}},\ \bibinfo {pages} {074103}
  (\bibinfo {year} {2017})}\BibitemShut {NoStop}%
\bibitem [{\citenamefont {Tribedi}, \citenamefont {Chakraborty},\ and\
  \citenamefont {Maitra}(2020)}]{tribedi2020formulation}%
  \BibitemOpen
  \bibfield  {author} {\bibinfo {author} {\bibfnamefont {S.}~\bibnamefont
  {Tribedi}}, \bibinfo {author} {\bibfnamefont {A.}~\bibnamefont
  {Chakraborty}}, \ and\ \bibinfo {author} {\bibfnamefont {R.}~\bibnamefont
  {Maitra}},\ }\href@noop {} {\bibfield  {journal} {\bibinfo  {journal}
  {Journal of Chemical Theory and Computation}\ }\textbf {\bibinfo {volume}
  {16}},\ \bibinfo {pages} {6317–6328} (\bibinfo {year} {2020})}\BibitemShut
  {NoStop}%
\bibitem [{\citenamefont {Feigenbaum}(1978)}]{feigenbaum1978quantitative}%
  \BibitemOpen
  \bibfield  {author} {\bibinfo {author} {\bibfnamefont {M.~J.}\ \bibnamefont
  {Feigenbaum}},\ }\href@noop {} {\bibfield  {journal} {\bibinfo  {journal}
  {Journal of statistical physics}\ }\textbf {\bibinfo {volume} {19}},\
  \bibinfo {pages} {25} (\bibinfo {year} {1978})}\BibitemShut {NoStop}%
\bibitem [{\citenamefont {Agarawal}, \citenamefont {Chakraborty},\ and\
  \citenamefont {Maitra}(2020)}]{agarawal2020stability}%
  \BibitemOpen
  \bibfield  {author} {\bibinfo {author} {\bibfnamefont {V.}~\bibnamefont
  {Agarawal}}, \bibinfo {author} {\bibfnamefont {A.}~\bibnamefont
  {Chakraborty}}, \ and\ \bibinfo {author} {\bibfnamefont {R.}~\bibnamefont
  {Maitra}},\ }\href@noop {} {\bibfield  {journal} {\bibinfo  {journal} {The
  Journal of Chemical Physics}\ }\textbf {\bibinfo {volume} {153}},\ \bibinfo
  {pages} {084113} (\bibinfo {year} {2020})}\BibitemShut {NoStop}%
\bibitem [{\citenamefont {Haken}(1989)}]{Haken_1989}%
  \BibitemOpen
  \bibfield  {author} {\bibinfo {author} {\bibfnamefont {H.}~\bibnamefont
  {Haken}},\ }\href {\doibase 10.1088/0034-4885/52/5/001} {\bibfield  {journal}
  {\bibinfo  {journal} {Rep. Prog. Phys.}\ }\textbf {\bibinfo {volume} {52}},\
  \bibinfo {pages} {515} (\bibinfo {year} {1989})}\BibitemShut {NoStop}%
\bibitem [{\citenamefont {Haken}\ and\ \citenamefont
  {Wunderlin}(1982)}]{haken1982slaving}%
  \BibitemOpen
  \bibfield  {author} {\bibinfo {author} {\bibfnamefont {H.}~\bibnamefont
  {Haken}}\ and\ \bibinfo {author} {\bibfnamefont {A.}~\bibnamefont
  {Wunderlin}},\ }\href@noop {} {\bibfield  {journal} {\bibinfo  {journal} {Z.
  Phys. B}\ }\textbf {\bibinfo {volume} {47}},\ \bibinfo {pages} {179}
  (\bibinfo {year} {1982})}\BibitemShut {NoStop}%
\bibitem [{\citenamefont {Haken}(1983)}]{Haken1983}%
  \BibitemOpen
  \bibfield  {author} {\bibinfo {author} {\bibfnamefont {H.}~\bibnamefont
  {Haken}},\ }\enquote {\bibinfo {title} {Nonlinear equations. the slaving
  principle},}\ in\ \href {\doibase 10.1007/978-3-642-45553-7_7} {\emph
  {\bibinfo {booktitle} {Advanced Synergetics: Instability Hierarchies of
  Self-Organizing Systems and Devices}}}\ (\bibinfo  {publisher} {Springer
  Berlin Heidelberg},\ \bibinfo {address} {Berlin, Heidelberg},\ \bibinfo
  {year} {1983})\ pp.\ \bibinfo {pages} {187--221}\BibitemShut {NoStop}%
\bibitem [{\citenamefont {Pedregosa}\ \emph {et~al.}(2011)\citenamefont
  {Pedregosa}, \citenamefont {Varoquaux}, \citenamefont {Gramfort},
  \citenamefont {Michel}, \citenamefont {Thirion}, \citenamefont {Grisel},
  \citenamefont {Blondel}, \citenamefont {Prettenhofer}, \citenamefont {Weiss},
  \citenamefont {Dubourg}, \citenamefont {Vanderplas}, \citenamefont {Passos},
  \citenamefont {Cournapeau}, \citenamefont {Brucher}, \citenamefont {Perrot},\
  and\ \citenamefont {Duchesnay}}]{scikit-learn}%
  \BibitemOpen
  \bibfield  {author} {\bibinfo {author} {\bibfnamefont {F.}~\bibnamefont
  {Pedregosa}}, \bibinfo {author} {\bibfnamefont {G.}~\bibnamefont
  {Varoquaux}}, \bibinfo {author} {\bibfnamefont {A.}~\bibnamefont {Gramfort}},
  \bibinfo {author} {\bibfnamefont {V.}~\bibnamefont {Michel}}, \bibinfo
  {author} {\bibfnamefont {B.}~\bibnamefont {Thirion}}, \bibinfo {author}
  {\bibfnamefont {O.}~\bibnamefont {Grisel}}, \bibinfo {author} {\bibfnamefont
  {M.}~\bibnamefont {Blondel}}, \bibinfo {author} {\bibfnamefont
  {P.}~\bibnamefont {Prettenhofer}}, \bibinfo {author} {\bibfnamefont
  {R.}~\bibnamefont {Weiss}}, \bibinfo {author} {\bibfnamefont
  {V.}~\bibnamefont {Dubourg}}, \bibinfo {author} {\bibfnamefont
  {J.}~\bibnamefont {Vanderplas}}, \bibinfo {author} {\bibfnamefont
  {A.}~\bibnamefont {Passos}}, \bibinfo {author} {\bibfnamefont
  {D.}~\bibnamefont {Cournapeau}}, \bibinfo {author} {\bibfnamefont
  {M.}~\bibnamefont {Brucher}}, \bibinfo {author} {\bibfnamefont
  {M.}~\bibnamefont {Perrot}}, \ and\ \bibinfo {author} {\bibfnamefont
  {E.}~\bibnamefont {Duchesnay}},\ }\href@noop {} {\bibfield  {journal}
  {\bibinfo  {journal} {Journal of Machine Learning Research}\ }\textbf
  {\bibinfo {volume} {12}},\ \bibinfo {pages} {2825} (\bibinfo {year}
  {2011})}\BibitemShut {NoStop}%
\bibitem [{\citenamefont {Agarawal}\ \emph {et~al.}(2021)\citenamefont
  {Agarawal}, \citenamefont {Roy}, \citenamefont {Chakraborty},\ and\
  \citenamefont {Maitra}}]{agarawal2021accelerating}%
  \BibitemOpen
  \bibfield  {author} {\bibinfo {author} {\bibfnamefont {V.}~\bibnamefont
  {Agarawal}}, \bibinfo {author} {\bibfnamefont {S.}~\bibnamefont {Roy}},
  \bibinfo {author} {\bibfnamefont {A.}~\bibnamefont {Chakraborty}}, \ and\
  \bibinfo {author} {\bibfnamefont {R.}~\bibnamefont {Maitra}},\ }\href@noop {}
  {\bibfield  {journal} {\bibinfo  {journal} {The Journal of Chemical Physics}\
  }\textbf {\bibinfo {volume} {154}},\ \bibinfo {pages} {044110} (\bibinfo
  {year} {2021})}\BibitemShut {NoStop}%
\bibitem [{\citenamefont {Haken}(2013)}]{haken2013synergetics}%
  \BibitemOpen
  \bibfield  {author} {\bibinfo {author} {\bibfnamefont {H.}~\bibnamefont
  {Haken}},\ }\href@noop {} {\emph {\bibinfo {title} {Synergetics: Introduction
  and advanced topics}}}\ (\bibinfo  {publisher} {Springer Science \& Business
  Media},\ \bibinfo {year} {2013})\BibitemShut {NoStop}%
\end{thebibliography}
    
\end{document}